# Physics as the science of the possible: Discovery in the age of Gödel

Dragutin T. Mihailović[*,(1)], Darko Kapor[*,(1)], Siniša Crvenković[*,(2)] and Anja Mihailovic[**]
University of Novi Sad, Novi Sad, Serbia
[*]Faculty of Sciences, [**]Faculty of Technical Sciences
[(1)]Department of Physics, [(2)] Department of Mathematics and Informatics

## Chapter 1

**Prolegomena**

This book represents a continuation, an elaboration, and possibly a clear explanation of the ideas which were expounded in the previous book *Time and Methods in Environmental Interfaces Modeling* (henceforth abbreviated as TM, Mihailović et al 2016). In that book as well as in whole of our published scientific work we were either implicitly or explicitly driven by a need to understand how the "space" between the human mind and observed physical reality is bridged. Here we use synonymously the terms "physical reality" and "reality" since the reality is all of physical existence, and concepts related to it as opposed to those products of our mind which remain on the level of mind. Relying on that book we add our new experiences in research in which physics plays a dominant role. To these experiences we attached some epistemological features as well as a view of physics through the optics of Gödel's Incompleteness Theorems (Gödel 1931). In the Prolegomena (Chapter 1) we consider some aspects of generality of physics (1.1 Generality of physics)

**1.1 Generality of physics**

According to many dictionaries physics is the branch of science concerned with the nature and properties of matter and energy. The subject of physics includes mechanics, heat, light and other radiation, sound, electricity, magnetism, and the structure of substance. In this definition the concept of the matter is used in the meaning of material substance that constitutes the observable universe and, together with energy (also appearing in the form of fields), forms the basis of all objective phenomena. *Inert matter* means all matter that is not a seed, including broken seeds, sterile florets, chaff, fungus bodies and stones. Remarkable part of the physical community takes it for granted that inertia is intrinsic to matter, although German physicist Ernst Mach had argued in 1872 that inertia has meaning only in reference to the matter in the universe. In empty space, you would not know that you were moving. On the other hand living *things* are highly organized, meaning they contain specialized, coordinated parts. All living organisms are made up of one or more cells, which are considered to be the fundamental units of life. Even unicellular organisms are complex! (There exists a radical difference between complicated object and complex one; thus, a chip is very complicated but fewer complexes compared to a cell). Multicellular organisms—such as humans—are made up of many cells.

    It appears that the pioneers in exploring these two worlds are physics and biology. It has been prevailing statement in science today that physics is the general, and hence, that biology is merely particular. This assertion in regard to physics is inconsequentially true. Rather, it could be

said that the aspiration of physics, or its dream, is to encompass all matter in nature in its all manifestations. However, organisms are part of a material nature and are involved in this encompass. Therefore, from such ideal perspectives biology is really a part of physics (when we say physics we mean *contemporary* physics). Contemporary physics, the physics we can find in the books and journals today, rests on assumptions that restrict it, keenly. For that reason, it is a very special science, applicable only to *very* special material systems; it is inherently inadequate to accommodate the phenomena at the heart of biology. No amount of sophistication within these limitations can compensate for the limitations themselves (Rosen 1991).

Erwin Schrödinger, the author of the book "What is life?", was one of the outstanding theoretical physicists of 20th century, perhaps of millennium. He had respectable knowledge in physics, in particular in statistical physics and thermodynamics. He viewed physics itself as the ultimate science of *material* nature, including of course those material systems we call organisms. He concluded that organisms were repositories of what he called *new physics*. Robert Rosen said that Erwin Schrödinger while permanently asserting the *universality* of contemporary physics, equally repeatedly pointed out (quite rightly) the utter failure of its laws to say anything significant about the biosphere and what is in it. What he was trying to say was stated a little later, perhaps even more vividly, by Albert Einstein. In a letter to Leo Szilard, Einstein said, "One can best feel in dealing with living things *how primitive physics still is*" (*emphasis added by* Robert Rosen). Consider, by contrast, the words of Jacques Monod, writing some three decades after the appearance of Schrödinger essay: "Biology is *marginal* because—the living world constituting but a tiny and very "special" part of the universe—it does not seem likely that the study of living things will ever uncover general laws applicable outside the biosphere). (*emphasis added* by Robert Rosen).

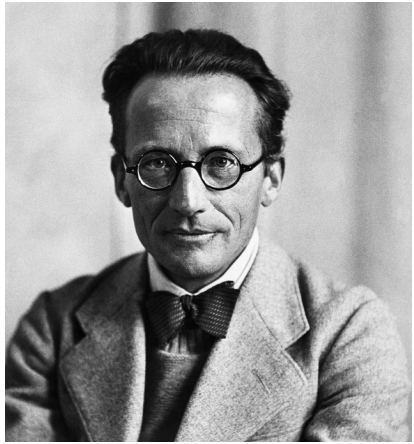

Erwin Schrödinger (1887-1961)

In this book, we will not cope with physics in the context of biology. When it comes to biology, then the works of Robert Rosen are the right address to understand the role of physics. We will cope with the breakthroughs of physics in biology and medicine where physical processes are obvious and clearly defined. So what is encompassed by biophysics? Since physics rightly "interferes" in many other segments embraced by the world of biosphere and civilization, we will deal with them. In that sense, maybe a right direction is to list the problems in the 21st century that will be in the focus of the world scientific community. One possible choice, without hierarchical order, is: (1) Whether we are alone in the Universe?; (2) Emission of harmful greenhouse gasses and climate change; (3) What is the consciousness?; (4) How to make decisions in an insecure world?; (5) Extension of both the maximum and average lifespan; (6) Whether the culture is characteristic only of people?; (7) Managing the Earth resources; (8) Importance of the Internet; (9) Use of the master cells in the future; (10) Importance of maintaining biodiversity; (11) The role of reengineering and climate change and (12) Importance of new vaccines.

What is the level of generality of physics or any other scientific or mathematical discipline? In order to answer this question, it is necessary to make an assessment, which is extremely difficult, if not impossible. This in fact rises a metaphoric question; a question *about* theory, not a question *within* the theory. We can say intuitively that the "level of generality" of a theory characterizes the class of situations with which the theory can cope, the class it can in principle accommodate. How, if at all, can something like that be measured? Actually, how?

It is illustrative, in this regard, to look at Number Theory comprising many conjectures, which no one has ever been able either to prove or produce counterexample (disprove). The rising question was: Is the Number Theory general enough, even in principle, to cope with situations that have arisen? The situation became more interesting when, working on undecibility in Number Theory, Kurt Gödel showed how to represent assertion *about* Number Theory *within* Number Theory (Gödel 1931). Considering the things in this way he was able to show that Number Theory is not complete or in other words: given any set of axioms for Number Theory, there are always propositions that are in some sense theorems but cannot be proved from these axioms.

If such a situation already exists in Number Theory then we can only imagine how difficult it is to ask a similar question about Physics. Stephen Hawking believed that Gödel's Incompleteness Theorems (Gödel 1931) makes the search for a *Theory of Everything* impossible. He reasoned that because there exist *mathematical results* that cannot be proven, there exist *physical results* that cannot be proven as well. The natural question is: Exactly how valid is his reasoning? Opinions on this view were and still are very divergent. We support the opinion that Stephen Hawking does not talk about results being unprovable in some abstract sense. He assumes that a *Theory of Everything* would be a particular finite set of rules, and he presents an argument that no such set of rules would be sufficient. He talks about "the final statement that it may not be possible to formulate a theory of the universe in a finite number of statements, which is reminiscent of Gödel's theorem" (Gonzalo, 2002).

Let us go back to physics. Above question is exactly question raised by reductionism in physics that is understood as a methodological reductionism which involves the attempt to reduce explanations to smaller constituents and to explain phenomena in terms of relations between more fundamental entities (Meyer-Ortmanns, 2015), connecting theories and knowledge levels *within physics*. Note that the problem with physical reductionism, at least as naively applied, is that: (i) it misses emergent properties of the system (reductionism says that emergent properties are nothing more than the sum of the reduced properties applied over a very large scale) and (ii) its misapplication in biology and other sciences (Longo, 2016). The question we talk about is an assertion, or conjecture, or belief, pertaining to the generality of contemporary physics itself. And indeed, the conjectures in physical world are not conjectures based on any *direct* evidence, for example, like Goldbach's Conjecture in Number Theory. It is rather *indirect* (circumstantial) evidence, insofar as evidence is adduced at all. In short, it rests on *faith*. This kind of faith was metaphorically described by the Serbian writer Borislav Pekić in the novel "Atlantida", where he says "The existence of spirits in principle does not contradict any law of physics. It is in contrast to the mind that civilization has modeled on empirical evidence for centuries."

However, the conjectures are based on general experience and in certain sense very limited since they are always possible subject to so called "black swan" effect. "Black swan" is an expression coined for the event (phenomenon) which is very rare, completely unexpected and completely unpredictable with large impact so that it can easily cause the decay of a complete scientific structure based on the conjecture which excluded (denied) its existence (See more in "Black Swan" by Nassim Nicholas Taleb).

At the end of this subchapter, we must stress that everything we said about physical systems is quite general, yet, later we shall make a clear distinction of the systems with infinite (or very large number of constituents) and the systems with interfaces compared to the rest of physical systems.